\documentclass[conference, 10pt]{IEEEtran}
\IEEEoverridecommandlockouts

\usepackage{cite}
\usepackage{amsmath,amssymb,amsfonts}
\usepackage{algorithm}
\usepackage{algorithmic}
\usepackage{graphicx}
\usepackage{epstopdf}
\usepackage{mathrsfs}
\usepackage{mathtools}
\usepackage{verbatim}
\usepackage{textcomp}
\usepackage{xcolor}

\DeclareMathOperator{\rect}{rect}
\DeclareMathOperator{\atan}{atan}

\newcommand{\lb}{{\ell}}

\def\BibTeX{{\rm B\kern-.05em{\sc i\kern-.025em b}\kern-.08em
    T\kern-.1667em\lower.7ex\hbox{E}\kern-.125emX}}
\begin{document}
\title{Characterizing the Narrowband Ambiguity Function of Frequency Modulated OFDM Waveforms
\thanks{This research was funded by the Naval Undersea Warfare Center's In-House Laboratory Independent Research (ILIR) program from the Office of Naval Research (ONR) under N0001424WX00177.}
}

\author{\IEEEauthorblockN{David A. Hague}
\IEEEauthorblockA{\textit{Naval Undersea Warfare Center} \\
1176 Howell St., Newport, RI, USA \\
david.a.hague.civ@us.navy.mil}
}

\maketitle

\begin{abstract}
This paper characterizes the spectral and Ambiguity Function (AF) properties of Frequency Modulated Orthogonal Frequency Division Multiplexing (FM-OFDM) waveforms employing Phase-Shift Keying (PSK).  The mainlobe structure of the AF is well approximated by the Ellipse of Ambiguity (EOA) model.  The EOA produces exact closed-form expressions that describe the AF mainlobe width in time-delay and Doppler as well as the degree of coupling between them.  These expressions yield three fundamental insights into the spectral and AF shapes of FM-OFDM waveforms employing PSK:  (1) these waveforms possess a fixed Root Mean Square (RMS) bandwidth for a given modulation index $h$, pulse-length $T$, and number of sub-carriers $L$, (2) they are roughly twice as spectrally efficient as Constant Envelope OFDM (CE-OFDM) waveforms with the same RMS bandwidth, and (3) for practical waveform design parameters will always possess a ``Thumbtack-Like'' AF shape.  Finally, this paper demonstrates a waveform optimization technique that minimizes the FM-OFDM waveform's Auto-Correlation Function (ACF) sidelobes using only a subset of the sub-carrier symbols as sensing ``pilot'' symbols leaving the rest for communications.  This pilot optimization technique is then demonstrated via two illustrative design examples.  
\end{abstract}

\begin{IEEEkeywords}
FM-OFDM, ISAC Waveform Design
\end{IEEEkeywords}

\section{Introduction}
Integrated Sensing and Communications (ISAC) techniques utilize the available electromagnetic spectrum and their respective system hardware to perform both sensing and communications functions simultaneously \cite{ISAC_Survey}.  Of the many design considerations for ISAC systems, novel waveform design promises to play an integral role in enabling this dual function capability.  ISAC waveform design is often divided into three categories \cite{ISAC_Waveform}.  Communication-Centric Waveform Design (CCWD) focuses on the communications aspects of waveform design with some consideration for their sensing abilities.  Sensing-Centric Waveform Design (SCWD) largely focuses on designing the waveforms for desirable sensing properties with some accomodation for communications.  Lastly, Joint Waveform Optimization and Design (JWOD) aims to jointly optimize the waveform to accomodate both the sensing and communication functions simultaneously to potentially provide a better tradeoff between the two.  All three design categories will require sophisticated waveform design techniques to realize the full potential ISAC systems promise. 

The majority of ISAC waveform design efforts have focused on Multi-Carrier (MC) waveforms \cite{visaMC} such as Orthogonal Frequency Division Multiplexing (OFDM) \cite{Sen_CE_OFDM} and the class of delay-Doppler communications waveforms \cite{Hong_Delay_Doppler_2022} including Orthogonal Time-Frequency Space (OTFS) modulation \cite{Calderbank_OTFS_I} and Orthogonal Delay-Doppler Division Multiplexing (ODDM) \cite{ODDM} modulation.  The primary drawback to these MC waveforms is that they possess a varying complex envelope resulting in a high Peak-to-Average Power Ratio (PAPR).  A higher PAPR requires input power backoff at the amplifier of the transmission system thus reducing overall energy efficiency of the transmitter.  Controlling the PAPR of ISAC waveforms remains an important research topic in the ISAC literature \cite{BazziI, BazziII}.  The CE-OFDM waveform is a constant envelope analogue of conventional OFDM where the symbol encoding is performed in the instantaneous phase domain \cite{Thompson_2, Chung_Comm}.  The CE-OFDM waveform encoding is essentially a form of Phase Modulation (PM) which guarantees a constant envelope.  This constant envelope property makes these waveforms better suited for efficient transmission on practical transmitter electronics than conventional MC waveforms \cite{Thompson_2}.  

More recently, Frequency Modulated OFDM (FM-OFDM) has been proposed as another constant envelope waveform for use in ISAC applications \cite{FM-OFDM, FM-OFDM-3}.  Simply stated, FM-OFDM implements OFDM encoding in the derivative of the waveform's instantaneous phase function (i.e, the frequency modulation function).  This encoding produces a waveform with higher spectral efficiency than CE-OFDM waveforms while retaining a constant modulus.  Most efforts in the literature on FM-OFDM have focused on CCWD applications such as defining the FM-OFDM waveform's communications receiver design \cite{FM-OFDM-2, FM-OFDM-3}, evaluating its channel estimation performance \cite{FM-OFDM-5}, and investigating its robustness to time-varying channels \cite{FM-OFDM-4}.  To the best of the author's knowledge, little is known about the FM-OFDM waveform's sensing characteristics.  A better understanding of these sensing characteristics could provide insight into how FM-OFDM waveforms could be utilized for SCWD and JWOD ISAC applications.

This paper takes a first step in fully mathematically characterizing the FM-OFDM waveform's Ambiguity Function (AF) shape.  This work utilizes a FM-OFDM waveform model that employs Phase-Shift Keying (PSK) to encode symbols and shows that the FM-OFDM waveform is mathematically very similar to the CE-OFDM waveform.  It is also a special case of the Multi-Tone Sinusoidal FM (MTSFM) waveform, an adaptive radar/sonar waveform described in \cite{Hague_AES, Hague_OCEANS_2022}.  Leveraging results from the efforts in \cite{Hague_AES, Hague_Felton_CE_OFDM_1, Hague_Felton_CE_OFDM_2} allows for fully characterizing the mainlobe structure of the FM-OFDM waveform's AF using the Ellipse of Ambiguity (EOA) model from the well established radar literature \cite{Cook, Rihaczek}.  These closed form expressions from the EOA model yield several insights into the spectral and AF characteristics of FM-OFDM waveforms employing PSK:  (1) these waveforms possess a fixed RMS bandwidth for a given modulation index $h$, pulse-length $T$, and number of sub-carriers $L$, (2) FM-OFDM waveforms are roughly twice as spectrally efficient as CE-OFDM waveforms with the same RMS bandwidth, and (3) for practical design parameters, these waveforms will always possess a ``Thumbtack-Like'' AF shape.  Finally, this paper demonstrates a waveform optimization technique that minimizes the FM-OFDM waveform's Auto-Correlation Function (ACF) sidelobes using only a subset $P \leq L$ of the sub-carrier symbols leaving the rest of the $Q = L-P$ sub-carrier symbols for communications.  These sensing symbols are loosely analogous to the pilot symbols used in MC waveforms.  Much like their MC counterpart, choice of the number of FM-OFDM pilot symbols introduces a design tradeoff between sensing waveform optimization performance and communications throughput.  This optimization technique is then demonstrated via two illustrative design examples. 

\section{The FM-OFDM Waveform Model}
\label{sec:sigModel}
This section describes the mainlobe and sidelobe structure of the AF and formally defines the FM-OFDM waveform model. 
\subsection{The Structure of the Ambiguity Function}
In general, a basebanded FM pulse waveform is expressed in continuous time as
\begin{equation}
s\left(t\right)=\frac{\rect\left(t/T\right)}{\sqrt{T}} e^{j\varphi\left(t\right)}
\label{eq:complexExpo}
\end{equation}
where $T$ is the waveform's pulse-length and $\varphi\left(t\right)$ is its phase modulation function. The time support is $|t| \leq T/2$, and $1/\sqrt{T}$ normalizes the signal energy to unity.  The waveform's frequency modulation function $m\left(t\right)$ maps its instantaneous frequency as a function of time and is expressed as
\begin{equation}
m\left(t\right) = \dfrac{1}{2\pi}\dfrac{\partial\varphi(t)}{\partial t}.
\label{eq:modFunc}
\end{equation}
The AF measures the response of the waveform's MF to its Doppler shifted versions and is defined as \cite{Cook, Rihaczek, Ricker}
\begin{IEEEeqnarray}{rCl}
\chi\left(\tau, \nu\right) = \int_{-\infty}^{\infty}s\left(t-\frac{\tau}{2}\right)s^*\left(t+\frac{\tau}{2}\right)e^{j2\pi \nu t} dt
\label{eq:AF}
\end{IEEEeqnarray}
where $\nu$ is the Doppler frequency shift.  Lastly, the ACF is the zero Doppler cut of the AF $R\left(\tau\right) = \chi\left(\tau, 0\right)$.  There are a series of design metrics that characterize a waveform's AF \& ACF shapes in terms of its mainlobe and sidelobe structure.  These metrics are described next.

\subsubsection{Mainlobe Structure}
\label{subsubsec:mainlobe}
The AF mainlobe structure determines a waveform's ability to estimate the range and Doppler of a single point target as well as its ability to resolve multiple targets in range and Doppler.  The AF mainlobe can be approximated by a second order Taylor series expansion \cite{Cook}.  The EOA is the contour of this AF mainlobe approximation at some fixed height $\xi$ and is always a coupled ellipse \cite{Cook, Rihaczek, Ricker}.  The EOA for the AF is expressed as \cite{Ricker}
\begin{equation}
1 - |\chi\left(\tau, \nu\right)|^2 = \xi = \beta_{rms}^2\tau^2 + 2\rho\tau\nu + \tau_{rms}^2\nu^2
\end{equation}
where $\beta_{rms}^2$ is the waveform's RMS bandwidth and determines time-delay (range) sensitivity, $\tau_{rms}^2$ is the RMS pulse-length which determines Doppler (range-rate) sensitivity, and $\rho$ is the Range-Doppler Coupling Factor (RDCF) for the AF mainlobe.  

The RMS bandwidth is expressed as \cite{Cook, Cohen}
\begin{align}
\beta_{rms}^2 &= \int_{-\infty}^{\infty}\left(f - \langle f \rangle\right)^2 |S\left(f\right)|^2 df \\ &= \int_{\Omega_t} | \dot{s}\left(t\right)|^2 dt - \left| \int_{\Omega_t} s\left(t \right) \dot{s}^*\left(t\right) dt \right| ^2 \\ &= \dfrac{1}{T}\int_{-T/2}^{T/2} \left[\dot{\varphi}\left( t \right)\right]^2 dt - \left| \dfrac{1}{T}\int_{-T/2}^{T/2} j \dot{\varphi}\left(t\right) dt \right| ^2
\label{eq:Brms}
\end{align}
where $\langle f \rangle$ is the waveform's spectral centroid, $S\left(f\right)$ is the waveform's Fourier transform, $\dot{s}\left(t\right)$ is the first time derivative of the waveform $s\left(t\right)$, $\Omega_t$ represents the region of support in time of the waveform, and $\dot{\varphi}\left(t\right)=2\pi m\left(t\right)$ is the first time derivative of the waveform's instantaneous phase.  Note that \eqref{eq:Brms} results from inserting \eqref{eq:complexExpo} into (6).  The RMS pulse-length term is expressed as 
\begin{equation}
\tau_{rms}^2 = 4\pi^2\int_{\Omega_t} \left(t-\langle t \rangle\right)^2 |s\left(t\right)|^2 dt = \dfrac{\pi^2T^2}{3}
\label{eq:Trms}
\end{equation}
where $\langle t \rangle$ is the first time moment of the the waveform $s\left(t\right)$.  Note that the rightmost expression in \eqref{eq:Trms} results from inserting \eqref{eq:complexExpo} in the leftmost expression in \eqref{eq:Trms}.  Finally, the RDCF is expressed as 
\begin{equation}
\rho = -2\pi \Im \Biggl\{\int_{\Omega_t} ts\left(t\right)\dot{s}^*\left(t\right) dt \Biggr\} = 2\pi  \int_{-T/2}^{T/2}  t \dot{\varphi}\left(t\right) dt
\label{eq:rho}
\end{equation}
where $\Im \{\}$ denotes the imaginary component of the integral.  The second expression in \eqref{eq:rho} results from inserting \eqref{eq:complexExpo} into the first expression in \eqref{eq:rho}.  

For the waveform model in \eqref{eq:complexExpo}, the RMS pulse-length $\tau_{rms}^2$ is solely dependent upon the pulse-length $T$ of the waveform.  The RMS bandwidth and RDCF EOA parameters are largely dependent upon the modulation function \eqref{eq:modFunc}.  If the waveform's modulation function is known, $\beta_{rms}^2$ and $\rho$ can be calculated in exact closed form which along with $\tau_{rms}^2$ provides full characterization of the waveform's AF mainlobe structure \cite{Cook, Rihaczek, Ricker}.  Furthermore, if the waveform's modulation function is composed of a discrete set of coefficients, those coefficients can be chosen to design waveforms with specific EOA characteristics thus directly shaping the AF mainlobe structure \cite{Hague_UACE}.

\subsubsection{Sidelobe Structure}
\label{subsubsec:sidelobe}
This paper will focus specifically on the sidelobe structure of the ACF.  The primary metric for evaluating and optimizing the ACF sidelobe structure is the Generalized Integrated Sidelobe Ratio (GISR) \cite{Blunt_PCFM_Grad} which evaluates the ratio of $\ell_p$-norms \cite{Prabhu_Babu_II} of both the sidelobe and mainlobe regions of the ACF expressed as
\begin{IEEEeqnarray}{rCl}
\text{GISR}~= \left(\dfrac{\int_{\Omega_{\tau}}\left|R\left(\tau\right)\right|^p d\tau}{\int_{-\Delta \tau}^{\Delta \tau}\left|R\left(\tau\right)\right|^p d\tau}\right)^{2/p}
\label{eq:GISR}
\end{IEEEeqnarray} 
where $p \geq 2$ is an integer, $\pm\Delta\tau$ represent the first nulls of the ACF thus establishing the ACF mainlobe width to be $2\Delta\tau$, and $\Omega_{\tau}$ is a region of time-delays excluding the mainlobe region $|\tau| \geq \Delta \tau$.  When $p=2$, the GISR becomes the standard ISR metric as described in the literature \cite{Blunt_PCFM_Grad}.  As $p\rightarrow\infty$, the integrals in \eqref{eq:GISR} approach the infinity norm $||\cdot||_{\infty}^2$, also known as the max-norm, and correspondingly the GISR approaches the Peak-to-Sidelobe Level Ratio (PSLR)\cite{Blunt_PCFM_Grad}.  However, from a waveform optimization perspective, the max-norm can produce a discontinuous objective function which prevents the efficient use of gradient-descent based waveform optimization methods.  Making $p$ large but finite approximates the PSLR metric while also resulting in a smooth objective function which can be traversed using gradient-descent based techniques.  These desirable properties of the GISR make it the appropriate design metric to reduce the ACF sidelobe levels of FM-OFDM waveforms.

\subsection{The FM-OFDM Waveform}
\label{subsec:FM-OFDM}
The FM-OFDM waveform is realized as a FM waveform whose instantaneous frequency function is a form of OFDM modulation
\begin{equation}
m\left(t\right) = \frac{2\pi h}{T}\sum_{\ell=1}^L a_\lb \cos\left(\frac{2\pi \ell t}{T}\right) + b_\lb \sin\left(\frac{2\pi \lb t}{T}\right)
\label{eq:fmofdm_1}
\end{equation}
where $L$ is the number of sub-carriers in $m\left(t\right)$, $h$ is the waveform's modulation index which controls its operational bandwidth, and $a_\lb$ and $b_\lb$ represent the in-phase and quadrature values of the $L$ communication symbols in $m\left(t\right)$.  Using the amplitude-phase representation for real-valued Fourier series results in the final form of the FM-OFDM's phase modulation function
\begin{equation}
m\left(t\right) = \frac{2\pi h}{T}\sum_{\ell=1}^L \vert \Gamma_\lb \vert \cos\left(\frac{2\pi \ell t}{T} + \phi_\lb\right)
\label{eq:fmofdm_2}
\end{equation}
where $\vert\Gamma_{\ell}\vert$ and $\phi_{\ell}$ are the FM-OFDM symbol magnitudes and phases respectively
\begin{equation}
\vert\Gamma_{\ell} \vert = \sqrt{a_{\ell}^2 + b_{\ell}^2},~\phi_{\ell} = \atan\left\{\dfrac{-b_\lb}{a_\lb} \right\}.
\label{eq:fmofdm_3}
\end{equation}
This paper will use M-ary PSK (M-PSK) to modulate the sub-carrier symbols and thus $\vert \Gamma_\lb \vert = 1$.  Finally, integrating \eqref{eq:fmofdm_2} and multiplying by $2\pi$ results in the FM-OFDM waveform's phase modulation function
\begin{equation} 
\varphi\left(t\right)=2\pi h\sum_{\ell=1}^L\frac{\vert\Gamma_{\ell}\vert}{\ell}\sin\left(\dfrac{2\pi \ell t}{T}+\phi_{\ell}\right)
\label{eq:fmofdm_4}
\end{equation}

It is interesting to note here that the FM-OFDM waveform is very closely related to the CE-OFDM waveform model as defined in \cite{Blunt_CEOFDM, Hague_Felton_CE_OFDM_1}.  The primary differences between these waveforms is that the FM-OFDM waveform's phase contains sine harmonics instead of cosine harmonics and that each symbol is attenuated by the harmonic index $\ell$.  Additionally, much like the CE-OFDM waveform, the FM-OFDM waveform's frequency modulation function as defined in \eqref{eq:fmofdm_1} is in fact a special case of the MTSFM waveform model described in \cite{Hague_AES}.  The MTSFM's modulation function is a more general version of \eqref{eq:fmofdm_1} where $a_{\ell}$ and $b_{\ell}$ are allowed to vary in magnitude rather than strictly lie on the unit circle to implement PSK.  This very observation aided the author's efforts in describing the sensing properties of CE-OFDM waveforms in \cite{Hague_Felton_CE_OFDM_1}.  Therefore, many of the same mathematical methods used to describe the MTSFM and CE-OFDM waveform models are directly applicable to the FM-OFDM waveform model.

\section{FM-OFDM AF Mainlobe Properties}
The EOA parameters of the FM-OFDM, derived in the Appendix, are expressed as
\begin{align}
\beta_{rms}^2 &= \dfrac{8\pi^4h^2L}{T^2}, \label{eq:Brms1}\\
\tau_{rms}^2 &= \dfrac{\pi^2T^2}{3}, \label{eq:Trms1}\\
\rho &= 4\pi^2h\sum_{\ell=1}^L\frac{ \left(-1\right)^{\ell+1}}{\ell}\sin\left(\phi_{\ell}\right). \label{eq:rho1}
\end{align}
The RDCF $\rho$ can be further modified to the normalized RDCF $\tilde{\rho}$ which behaves like a correlation coefficient and takes on the values $-1 \leq \rho \leq 1$ \cite{Cook, Cohen} expressed as
\begin{equation}
\tilde{\rho} = \dfrac{\rho}{\tau_{rms}\beta_{rms}} = \sqrt{\frac{6}{L}}\left(\frac{1}{\pi}\right)\sum_{\ell=1}^L \dfrac{\left(-1\right)^{\ell + 1}}{\ell}\sin\phi_\lb.
\label{eq:rho2}
\end{equation}
The results shown in \eqref{eq:Brms1}-\eqref{eq:rho2} provide three fundamental insights into the spectral and AF characteristics of the FM-OFDM waveform.  The first is that the expression for $\beta_{rms}^2$ in \eqref{eq:Brms1} is an exact closed-form expression that stays fixed for all real values of $h$, $T$, and $L$ regardless of the PSK sequence encoded into the waveform.  The other two fundamental insights are subsequently described below.

\begin{figure}[ht]
\centering
\includegraphics[width=0.5\textwidth]{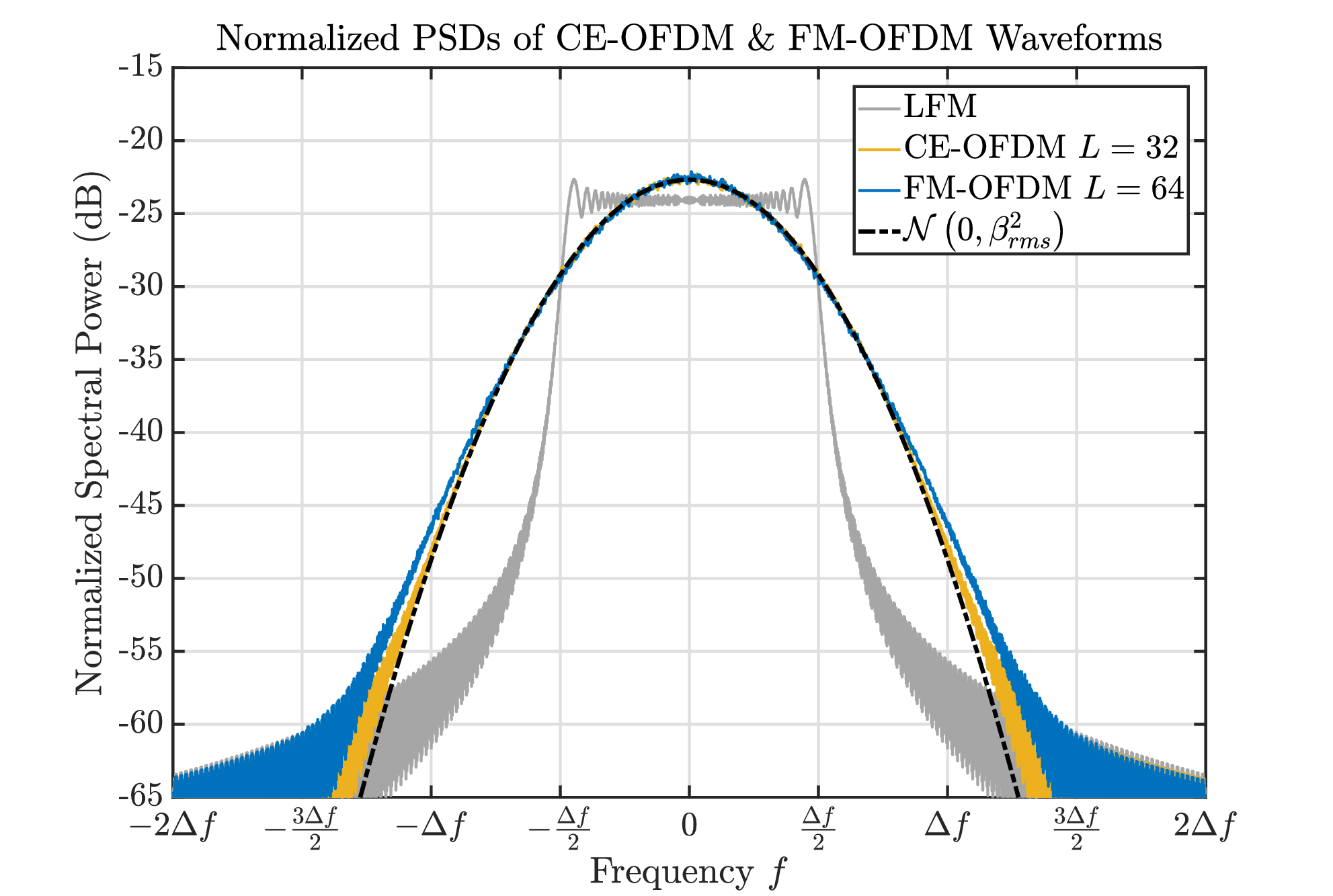}
\caption{Estimated PSDs of CE-OFDM and FM-OFDM waveforms with Gaussian PSD fit.  Also shown is the spectrum of the LFM with the same TBP.  The FM-OFDM waveform possesses essentially the same PSD shape while also containing twice as many sub-carriers $L$ as the CE-OFDM waveform for the same RMS bandwidth.  This suggests that FM-OFDM is roughly twice as spectrally efficient as CE-OFDM.}
\label{fig:FM-OFDM_1}
\end{figure}

The RMS bandwidth in \eqref{eq:Brms1} not only provides a measure of the FM-OFDM waveform's time-delay (range) resolution, but it also provides a measure of the waveform's spectral spread.  This is particularly useful for ISAC waveforms in that it helps to determine its spectral efficiency.  As was described in \cite{FM-OFDM, FM-OFDM-4}, the FM-OFDM's Power Spectral Density (PSD) is approximately Gaussian shaped.  It therefore stands to reason that  a zero-mean Gaussian function with variance $\beta_{rms}^2$ should be an accurate fit to the FM-OFDM waveform's PSD.  This is illustrated in Figure \ref{fig:FM-OFDM_1} which shows a PSD estimate of the FM-OFDM waveform using Welch's method derived from 1000 waveforms each with $L=64$ sub-carriers with $M_{\text{PSK}}=8$ (i.e, 3-bits per symbol) and whose RMS bandwidths are chosen using \eqref{eq:Brms1} to possess the same RMS bandwidth as a Linear FM (LFM) waveform with a Time-Bandwidth Product (TBP) of 256.  The modulation index is $2\pi h = 13.03$ and can be found by setting \eqref{eq:Brms1} to the RMS bandwidth of the LFM waveform $8\pi^4h^2L/T^2 = \pi^2\Delta f^2/3$ where $\pi^2\Delta f^2/3$ is the LFM waveform's RMS bandwidth and $\Delta f$ is its swept bandwidth.  Solving for $2\pi h$ yields $2\pi h = T\Delta f/\sqrt{6L}$.  Also shown is the estimated PSD of the CE-OFDM waveform also using Welch's method derived from 1000 waveforms each with $L=32$ sub-carriers and modulation index $2\pi h =0.98$ whose RMS bandwidth is also equal to that of the LFM waveform \cite{Hague_Felton_CE_OFDM_1}.  Note that making $L\geq32$ for the CE-OFDM waveform while maintaining the same RMS bandwidth would reduce its modulation index to the point that its PSD substantially departs from the Gaussian shape and thus $L=32$ represents the upper bound on $L$ for this TBP \cite{Thompson_2}.  From the figure, it is clear that both the FM-OFDM and CE-OFDM waveforms possess nearly identical PSD shapes and consequently the same RMS bandwidth approximately.  Both waveforms' PSDs are well modeled by a Gaussian function whose variance is the RMS bandwidth.  However, the FM-OFDM waveform is able to achieve this while employing double the number of sub-carriers.  Similar results hold for realizations of both waveforms for other TBP values.  This suggests that FM-OFDM is roughly twice as spectrally efficient as CE-OFDM.

\begin{figure}[ht]
\centering
\includegraphics[width=0.5\textwidth]{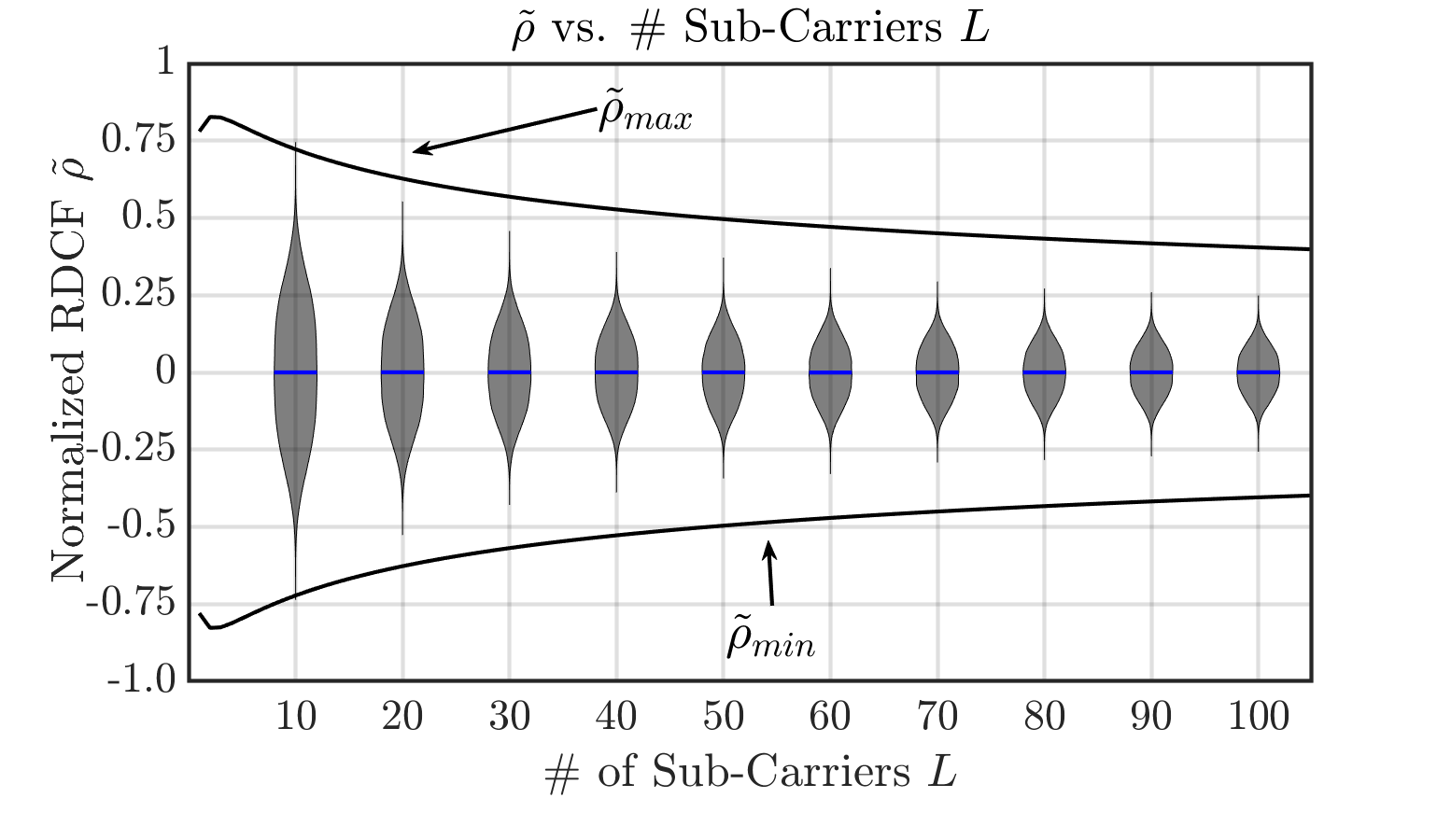}
\caption{Violin plots of  $\tilde{\rho}$ for FM-OFDM waveforms as a function of the number of sub-carriers $L$.  For practical values of $L$, the FM-OFDM waveform will always possess a "Thumbtack-Like" AF shape.}
\label{fig:FM-OFDM_2}
\end{figure}

The normalized RDCF $\tilde{\rho}$ is a function of the number of sub-carriers as well as their respective PSK values $\phi_{\ell}$.  This raises the intriguing question of whether the choice of PSK sequence has a substantial impact on the RDCF.  To investigate this, Figure \ref{fig:FM-OFDM_2} displays violin plots of the values of $\tilde{\rho}$ for 100,000 realizations of PSK sequences with $M_{\text{PSK}}=8$ for a varying number of sub-carriers $L$ going from 10 to 100 in increments of 10.  The blue lines on the violin plots show the median of $\tilde{\rho}$ and the shapes or ``violins'' give a pictorial representation of the distribution of $\tilde{\rho}$.  Also shown are the maximum and minimum values attainable for $\tilde{\rho}$.  The maximum value is achieved when the $\left(-1\right)^{\ell+1}\sin\left(\phi_{\ell}\right)$ term in \eqref{eq:rho2} is $+1$ for all $\ell$ and occurs when $\phi_{\ell}$ takes the form
\begin{equation}  \phi_{\ell}= \left\{
\begin{array}{ll}
      ~~\frac{\pi}{2},  &\ell~\text{odd} \\
      
      -\frac{\pi}{2},  &\ell~\text{even}. \\
\end{array} 
\right.
\label{eq:rho_min}
\end{equation}
Using \eqref{eq:rho2}, $\tilde{\rho}_{\text{max}}$ is expressed as
\begin{equation}
\tilde{\rho}_{\text{max}} = \sqrt{\dfrac{6}{L}}\left(\dfrac{1}{\pi}\right)\sum_{\ell=1}^L \dfrac{1}{\ell}=\sqrt{\dfrac{6}{L}}\left(\dfrac{1}{\pi}\right)H_L
\label{eq:rho_7}
\end{equation}
where $H_L$ is the Harmonic number.  Correspondingly, flipping the sequence in \eqref{eq:rho_min} produces the minimum of $\tilde{\rho}$ which is simply the negative of \eqref{eq:rho_7}.  From the figure, it's clear that the median value for $\tilde{\rho}$ is very close to zero and that as $L$ increases, the distribution of $\tilde{\rho}$ becomes increasingly concentrated near zero.  Since most FM-OFDM waveform realizations will possess 10's of sub-carriers, it stands to reason that FM-OFDM waveforms with practical design parameters will possess nearly zero range-Doppler coupling in their AF mainlobe and will attain a ``Thumbtack-Like'' AF shape.      

\section{Controlling the ACF Sidelobes of FM-OFDM Waveforms}
\label{sec:sidelobes}

Controlling the ACF sidelobes of the FM-OFDM waveform requires defining an optimization routine that minimizes the GISR metric over the FM-OFDM's symbol phases $\phi_\lb$.  Previous efforts in \cite{Hague_Felton_CE_OFDM_2} demonstrated a numerical optimization algorithm that reduced the ACF sidelobes of the CE-OFDM waveform.  The algorithm arrives at a local minimum of the GISR objective function by letting the M-PSK coefficients lie on the continuum of points on the unit circle rather than discrete points on the unit circle as required by the M-PSK constellation.  By doing this, the optimization routine produced a CE-OFDM waveform that was purely optimized for sensing removing all communications throughput.  

This paper develops an optimization routine that minimizes the FM-OFDM waveform's ACF sidelobes over a subset $\mathcal{S}$ of $P \leq L$ sub-carrier symbols denoted as $\phi_{\ell \in \mathcal{S}}$.  The remaining $Q=L-P$ set of sub-carrier symbols $\mathcal{C}$ are reserved for communications and are denoted as $\phi_{\ell \in \mathcal{C}}$.  The FM-OFDM model can be represented in terms of these two sets of sub-carriers as
\begin{equation}
s\left(t\right) = \dfrac{\rect\left(t/T\right)}{\sqrt{T}}e^{\varphi_{\ell\in\mathcal{C}}\left(t\right)}e^{\varphi_{\ell\in\mathcal{S}}\left(t\right)}
\end{equation}
where $\varphi_{\ell \in \mathcal{C}}\left(t\right)$ and $\varphi_{\ell \in \mathcal{S}}\left(t\right)$ are the instantaneous phase functions resulting from the communications and sensing symbols $\phi_{\ell \in \mathcal{C}}$ and $\phi_{\ell \in \mathcal{S}}$ respectively.  This modification is loosely analogous to the use of pilot symbols in OFDM which are typically used for channel characterization in communications applications \cite{visaMC}.  A similar method was proposed  for Affine Frequency Division Multiplexing (AFDM) in \cite{AFDM_AF}.  In this case, the FM-OFDM pilot symbols can be used to optimize the sensing characteristics of the waveform.  Modifying the values for $P$ and $Q$ in this pilot optimization technique introduces a design tradeoff between refining the sensing characteristics and communications throughput respectively. 

\begin{figure}[ht]
\centering
\includegraphics[width=0.5\textwidth]{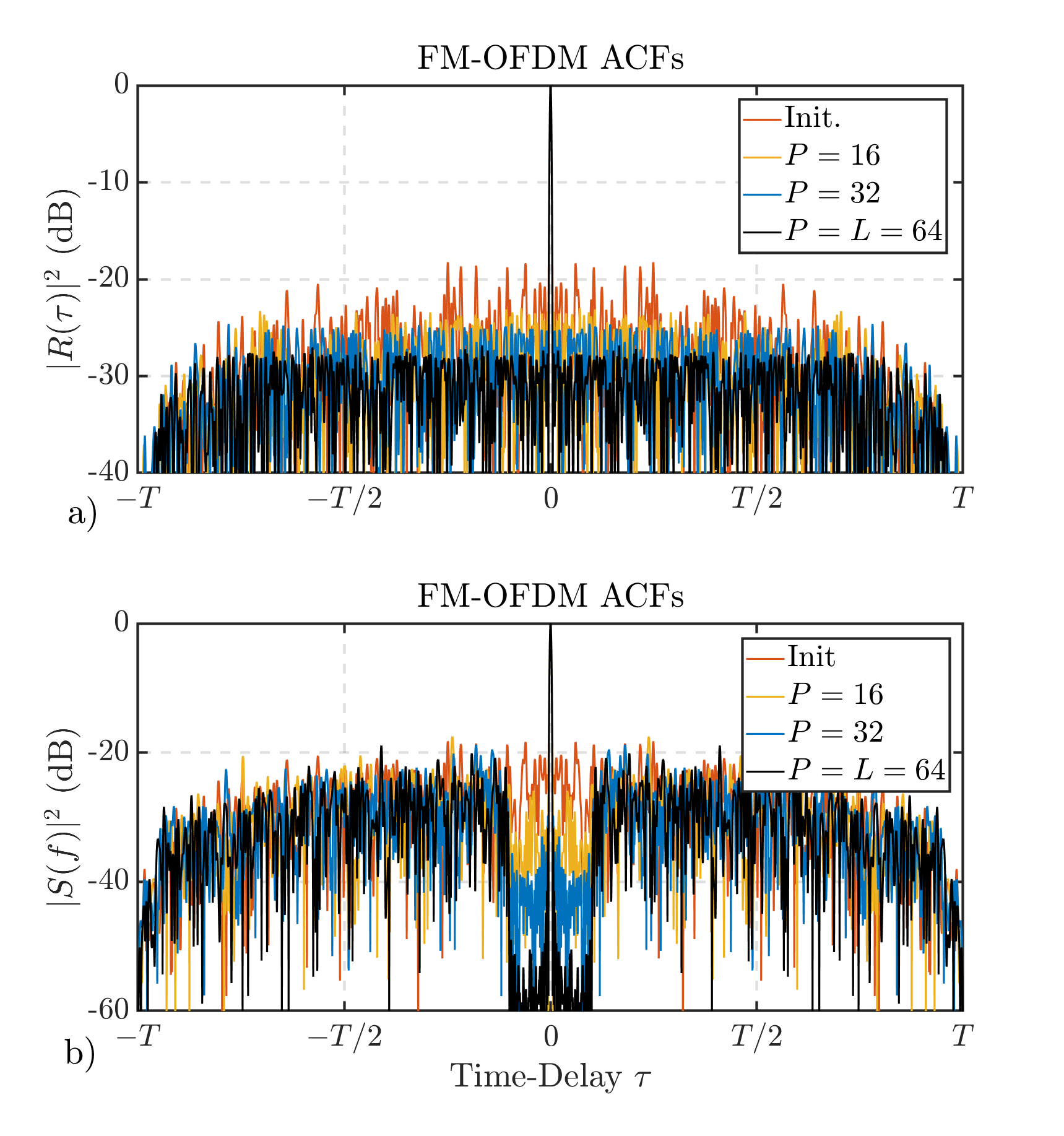}
\caption{FM-OFDM waveform ACFs optimized (a) over all time-delays and (b) over a sub-region of time-delays with a varying number of sensing symbols $P$.  A larger $P$ introduces greater degrees of freedom to the optimization problem and reduced the resulting waveform's ACF sidelobes in the prescribed region of time-delays.  This enhanced sensing performance comes at the cost of reduced communications throughput.}
\label{fig:FM-OFDM_4}
\end{figure}

This tradeoff is demonstrated by minimizing the ACF sidelobes over sub-regions in time-delay using the GISR metric.  Formally, the optimization problem can be stated as 
\begin{equation}
\underset{\phi_{\ell \in \mathcal{S}}}{\text{min}}~\text{GISR}\{p, \Delta\tau, \Omega_{\tau}, \phi_{\ell \in \mathcal{S}}\}.
\label{eq:Problem_1}
\end{equation}
The optimization problem in \eqref{eq:Problem_1} is solved using MATLAB's \emph{fminunc()} function.  It is important to note that this routine will only find a local minimum and is not streamlined for computational efficiency.  Figure \ref{fig:FM-OFDM_4} shows the resulting optimized waveforms with varying $P$ for two FM-OFDM waveform optimization problems.  The first optimizes the ACF sidelobes over all time-delays outside the mainlobe and the second over the region $\Omega_{\tau} \in \Delta\tau \leq |\tau| \leq  0.1T$.  The initial waveform passed to the optimization routine is composed of $L=64$ sub-carriers, a modulation index of $2\pi h = 13.03$, and a randomly initialized PSK sequence $\phi_{\ell}$.  This waveform possesses the same RMS bandwidth as an LFM waveform with a TBP of 256 just like those shown in Figure \ref{fig:FM-OFDM_1}.  Both optimization problems were run using $P = 16, 32, \&~64$.  As expected, a larger $P$ introduces more degrees of freedom in the optimization problem and this produced waveforms with noticeably lower ACF sidelobes.  However, these enhanced sensing characteristics came at the cost of reduced communications throughput highlighting the design tradeoff using this pilot-based waveform optimization technique.

\section{Conclusion}
\label{sec:Conclusion}
This paper derived several exact closed form expressions that characterize the spectral and AF shapes of FM-OFDM waveforms employing PSK encoding.  These results show that FM-OFDM waveforms are more spectrally efficient than CE-OFDM waveforms and almost always possess a "Thumbtack-Like" AF.  This paper also demonstrated a waveform optimization technique that minimizes the FM-OFDM waveform's Auto-Correlation Function (ACF) sidelobes using only a subset of the sub-carrier symbols as sensing ``pilot'' symbols leaving the rest for communications.  The illustrative design examples demonstrate that this pilot optimization technique introduces a design tradeoff between refining the sensing characteristics and communications throughput of the waveform. The results of this paper take a preliminary step in describing how FM-OFDM waveforms can potentially be optimized for SCWD and JWOD waveform design applications.  
\appendix[Deriving the FM-OFDM's EOA Parameters]
The Appendix derives the RMS bandwidth and RDCF of the rectangularly windowed FM-OFDM waveform.
\subsection{RMS Bandwidth}
\label{subsec:rms1}
The RMS bandwidth $\beta_{rms}^2$ is calculated via \eqref{eq:Brms}.  Inserting $2\pi m\left(t\right)$ into \eqref{eq:Brms} using \eqref{eq:fmofdm_2} results in the expression

\begin{multline}
 \beta_{rms}^2 = \dfrac{8 \pi^4 h^2}{T^2} \left(\dfrac{2}{T}\right)\int_{-T/2}^{T/2} \left[\sum_{\ell=1}^L a_{\ell}\cos\left(\dfrac{2\pi \ell t}{T} \right) + \right. \\ \left. b_{\ell}\sin\left(\dfrac{2\pi \ell t}{T} \right)  \right]^2dt.
\label{eq:brms_3}
\end{multline}
The scaling factor $2/T$ times the integral in \eqref{eq:brms_3} can be simplified by utilizing the Parseval's theorem for Fourier series resulting in the expression
\begin{IEEEeqnarray}{rCl}
\sum_{\ell=1}^L \left(a_{\ell}^2 + b_{\ell}^2\right) = \sum_{\ell=1}^L |\Gamma_{\ell}|^2 = L
\label{eq:brms_4}
\end{IEEEeqnarray}
which produces the final result
\begin{equation}
\beta_{rms}^2 = \dfrac{8\pi^4h^2 L}{T^2}.
\label{eq:brms_5}
\end{equation}

\subsection{Range-Doppler Coupling Factor}
\label{subsec:RDCF}
Inserting $2\pi m\left(t\right)$ into \eqref{eq:rho} using \eqref{eq:fmofdm_2} results in the expression
\begin{IEEEeqnarray}{rCl}
\rho = \dfrac{8\pi^3h}{T^2} \sum_{\ell=1}^L\left(\frac{1}{\ell}\right)\int_{-T/2}^{T/2} t \cos \left(\frac{2 \pi \ell t}{T}+\phi_{\ell}\right) dt.
\label{eq:rho_2}
\end{IEEEeqnarray}
Using the trigonometric identity $\cos\left(a\pm b\right) = \cos a \cos b \mp \sin a \sin b$, \eqref{eq:rho_2} expands to
\begin{multline}
\rho = -\dfrac{8\pi^3h}{T^2} \sum_{\ell=1}^L\left(\frac{1}{\ell}\right) \left[\cos\left(\phi_{\ell}\right) \int_{-T/2}^{T/2}t\cos\left(\dfrac{2 \pi \ell t}{T}\right) +  \right.  \\ \left. \sin\left(\phi_{\ell}\right) \int_{-T/2}^{T/2}t\sin\left(\dfrac{2 \pi \ell t}{T}\right) \right].
\label{eq:rho_3}
\end{multline}
The first integral evaluates to zero since the integrand is an odd-symmetric function evaluated over an even-symmetric interva.  Using integration by parts, the second integral in \eqref{eq:rho_3} evaluates to 
\begin{equation}
\dfrac{T^2}{2\pi \ell}\left(-1\right)^{\ell+1}\sin\phi_{\ell}.
\label{eq:rho_4}
\end{equation}
Inserting the result in \eqref{eq:rho_4} back into \eqref{eq:rho_3} along with $\vert \Gamma_{\ell} \vert = 1$ results in the RDCF $\rho$ for the FM-OFDM waveform
\begin{equation}
\rho = 4\pi^2 h \sum_{\ell=1}^L  \frac{\left(-1\right)^{\ell+1}}{\ell}\sin\left(\phi_{\ell}\right).
\label{eq:rho_5}
\end{equation}
Finally, the normalized RDCF $\tilde{\rho}$ is expressed as
\begin{align}
\tilde{\rho} = \dfrac{\rho}{\tau_{rms}\beta_{rms}} &= \dfrac{\sqrt{3}}{\pi T}\dfrac{T}{\sqrt{8}\pi^2 h \sqrt{L}}4\pi^2 h \sum_{\ell=1}^L  \frac{\left(-1\right)^{\ell+1}}{\ell}\sin\left(\phi_{\ell}\right) \IEEEnonumber \\ &= \sqrt{\frac{6}{L}}\left(\frac{1}{\pi}\right)\sum_\lb^L \dfrac{\left(-1\right)^{\ell + 1}}{\ell}\sin\phi_\lb.
\label{eq:rho_6}
\end{align}

\bibliographystyle{IEEEtran}
\bibliography{Hague_IEEE_ISAC_Workshop_2026}

@book{Cohen,
  title={Time-frequency Analysis},
  author={Cohen, L.},
  isbn={9780135945322},
  lccn={94039843},
  series={Electrical engineering signal processing},
  year={1995},
  publisher={Prentice Hall PTR}
}

@book{Cook,
  title={Radar signals: an introduction to theory and application},
  author={Cook, C.E. and Bernfeld, M.},
  lccn={68030803},
  series={Electrical science series},
  url={},
  year={1967},
  publisher={Academic Press}
}

@book{Ricker,
  title={Echo Signal Processing},
  author={D. W. Ricker},
  lccn={68025217},
  url={},
  year={2003},
  publisher={Kluwer}
}

@book{Rihaczek,
  title={Principles of high-resolution radar},
  author={Rihaczek, A.W.},
  lccn={68025217},
  url={},
  year={1969},
  publisher={McGraw-Hill}
}

@ARTICLE{ISAC_Survey,
  author={Liu, Fan and Cui, Yuanhao and Masouros, Christos and Xu, Jie and Han, Tony Xiao and Eldar, Yonina C. and Buzzi, Stefano},
  journal={IEEE Journal on Selected Areas in Communications}, 
  title={Integrated Sensing and Communications: Toward Dual-Functional Wireless Networks for 6G and Beyond}, 
  year={2022},
  volume={40},
  number={6},
  pages={1728-1767},
  doi={10.1109/JSAC.2022.3156632}}

@ARTICLE{ISAC_Waveform,
  author={Zhou, Wenxing and Zhang, Ruoyu and Chen, Guangyi and Wu, Wen},
  journal={IEEE Open Journal of the Communications Society}, 
  title={Integrated Sensing and Communication Waveform Design: A Survey}, 
  year={2022},
  volume={3},
  number={},
  pages={1930-1949},
  doi={10.1109/OJCOMS.2022.3215683}}

@ARTICLE{visaMC,
  author={Koivunen, Visa and Keskin, Musa Furkan and Wymeersch, Henk and Valkama, Mikko and González-Prelcic, Nuria},
  journal={IEEE Signal Processing Magazine}, 
  title={Multicarrier ISAC: Advances in waveform design, signal processing, and learning under nonidealities}, 
  year={2024},
  volume={41},
  number={5},
  pages={17-30},
  doi={10.1109/MSP.2024.3420492}}

@INPROCEEDINGS{BazziI,
  author={Bazzi, Ahmad and Chafii, Marwa},
  booktitle={ICC 2025 - IEEE International Conference on Communications}, 
  title={Designing Waveforms with Adjustable PAPR for Integrated Sensing and Communication}, 
  year={2025},
  volume={},
  number={},
  pages={6633-6638},
  doi={10.1109/ICC52391.2025.11161464}}

@ARTICLE{BazziII,
  author={Wang, Dexin and Bazzi, Ahmad and Chafii, Marwa},
  journal={IEEE Journal on Selected Areas in Information Theory}, 
  title={DRIP: A Versatile Family of Space–Time ISAC Discrete-Time Sequences}, 
  year={2026},
  volume={7},
  number={},
  pages={205-220},
  doi={10.1109/JSAIT.2026.3694721}}

@book{Hong_Delay_Doppler_2022,
	title={Delay-Doppler Communications: Principles and Applications},
	author={Hong, Yi and Thaj, Tharaj and Viterbo, Emanuele},
	year={2022},
	publisher={Academic Press}
}

@INPROCEEDINGS{Calderbank_OTFS_I,
	author={Hadani, R. and Rakib, S. and Tsatsanis, M. and Monk, A. and Goldsmith, A. J. and Molisch, A. F. and Calderbank, R.},
	booktitle={2017 IEEE Wireless Communications and Networking Conference (WCNC)}, 
	title={Orthogonal Time Frequency Space Modulation}, 
	year={2017},
	volume={},
	number={},
	pages={1-6},
	doi={10.1109/WCNC.2017.7925924}}

@ARTICLE{ODDM,
	author={Lin, Hai and Yuan, Jinhong},
	journal={IEEE Transactions on Wireless Communications}, 
	title={Orthogonal Delay-Doppler Division Multiplexing Modulation}, 
	year={2022},
	volume={21},
	number={12},
	pages={11024-11037},
	doi={10.1109/TWC.2022.3188776}}

@ARTICLE{Thompson_2,
  author={Thompson, Steve C. and Ahmed, Ahsen U. and Proakis, John G. and Zeidler, James R. and Geile, Michael J.},
  journal={IEEE Transactions on Communications}, 
  title={Constant Envelope OFDM}, 
  year={2008},
  volume={56},
  number={8},
  pages={1300-1312},
  doi={10.1109/TCOMM.2008.070043}}

@INPROCEEDINGS{Chung_Comm,
  author={Char-Dir Chung and Shih-Ming Cho},
  booktitle={Fifth Asia-Pacific Conference on ... and Fourth Optoelectronics and Communications Conference on Communications,}, 
  title={Constant-envelope orthogonal frequency division multiplexing modulation}, 
  year={1999},
  volume={1},
  number={},
  pages={629-632 vol.1},
  doi={10.1109/APCC.1999.824966}}

@ARTICLE{FM-OFDM,
	author={Lorca Hernando, Javier and Armada, Ana García},
	journal={IEEE Transactions on Communications}, 
	title={Frequency-Modulated OFDM: A New Waveform for High-Mobility Wireless Communications}, 
	year={2023},
	volume={71},
	number={1},
	pages={540-552},
	doi={10.1109/TCOMM.2022.3224212}}

@article{FM-OFDM-2,
  title={FM OFDM Unifying High Mobility Communications and Sensing},
  author={Bouziane, Amir and Arslan, Huseyin},
  journal={arXiv preprint arXiv:2508.16107},
  year={2025}
}

@misc{FM-OFDM-3,
      title={Constant-Envelope ISAC via FM-OFDM: Analytical Framework and Receiver Design}, 
      author={Bouziane, Amir and Arslan, Huseyin},
      year={2026},
      eprint={2508.16107},
      archivePrefix={arXiv},
      primaryClass={eess.SP},
      url={https://arxiv.org/abs/2508.16107}, 
}

@ARTICLE{FM-OFDM-4,
  author={Romano, Gianmarco and Di Gennaro, Giovanni and Buonanno, Amedeo and Buzzi, Stefano and Palmieri, Francesco A. N.},
  journal={IEEE Open Journal of the Communications Society}, 
  title={Investigating Frequency Modulated OFDM (FM-OFDM): Receiver Design and Analysis Over Time-Varying Channels}, 
  year={2025},
  volume={6},
  number={},
  pages={3760-3778},
  doi={10.1109/OJCOMS.2025.3559706}}

@INPROCEEDINGS{FM-OFDM-5,
  author={Méndez-Monsanto, Lianet and Caus, Màrius and Shaat, Musbah and Pérez-Neira, Ana I. and García Armada, Ana},
  booktitle={2024 IEEE Globecom Workshops (GC Wkshps)}, 
  title={Adaptive FM-OFDM for 6G Non-Terrestrial Networks}, 
  year={2024},
  volume={},
  number={},
  pages={1-6},
  doi={10.1109/GCWkshp64532.2024.11100841}}

@ARTICLE{Sen_CE_OFDM,
  author={Sen, Satyabrata and Nehorai, Arye},
  journal={IEEE Transactions on Signal Processing}, 
  title={Adaptive OFDM Radar for Target Detection in Multipath Scenarios}, 
  year={2011},
  volume={59},
  number={1},
  pages={78-90},
  doi={10.1109/TSP.2010.2086448}}

@INPROCEEDINGS{Blunt_CEOFDM,	
	author={Biehl, Erik R. and Mohr, Charles A. and Ravenscroft, Brandon and Blunt, Shannon D.},
	booktitle={2020 IEEE Radar Conference (RadarConf20)}, 
	title={Assessment of Constant Envelope OFDM as a Class of Random FM Radar Waveforms}, 
	year={2020},
	volume={},
	number={},
	pages={1-6},
	doi={10.1109/RadarConf2043947.2020.9266427}}

@ARTICLE{Hague_AES,
author={D. A. {Hague}},
journal={IEEE Transactions on Aerospace and Electronic Systems}, 
title={Adaptive Transmit Waveform Design Using Multitone Sinusoidal Frequency Modulation}, 
year={2021},
volume={57},
number={2},
pages={1274-1287},
doi={10.1109/TAES.2020.3046086}}

@INPROCEEDINGS{Hague_OCEANS_2022,
  author={Hague, David A.},
  booktitle={OCEANS 2022, Hampton Roads}, 
  title={Adaptive Comb Waveform Design for Reverberation Suppression}, 
  year={2022},
  volume={},
  number={},
  pages={1-8},
  doi={10.1109/OCEANS47191.2022.9977391}}

@INPROCEEDINGS{Hague_Felton_CE_OFDM_1,
  author={Felton, David G. and Hague, David A.},
  booktitle={2023 IEEE Radar Conference (RadarConf23)}, 
  title={Characterizing the Ambiguity Function of Constant-Envelope OFDM Waveforms}, 
  year={2023},
  volume={},
  number={},
  pages={1-6}}

@INPROCEEDINGS{Hague_Felton_CE_OFDM_2,
  author={Felton, David G. and Hague, David A.},
  booktitle={2023 IEEE Radar Conference (RadarConf23)}, 
  title={Gradient-Descent Based Optimization of Constant Envelope OFDM Waveforms}, 
  year={2023},
  volume={},
  number={},
  pages={1-6}}

@article{Hague_UACE,
  title={Characterizing the narrowband ambiguity function of multi-tone sinusoidal frequency modulated waveforms},
  author={Hague, David A},
  journal={arXiv preprint arXiv:2312.05188},
  year={2023}
}

@misc{Prabhu_Babu_II,
      title={Design of MIMO Radar Waveforms based on lp-Norm Criteria}, 
      author={E. Raei and M. Alaee-Kerahroodi and P. Babu and M. R. {Bhavani Shankar}},
      year={2021},
      eprint={2104.03303},
      archivePrefix={arXiv},
      primaryClass={eess.SP}
}

@ARTICLE{Blunt_PCFM_Grad,
  	author={Mohr, Charles A. and McCormick, Patrick M. and Topliff, Charles A. and Blunt, Shannon D. and Baden, J. Michael},
  	journal={IEEE Transactions on Aerospace and Electronic Systems}, 
  	title={Gradient-Based Optimization of PCFM Radar Waveforms}, 
  	year={2021},
  	volume={57},
  	number={2},
  	pages={935-956},
  	doi={10.1109/TAES.2020.3037403}}

@misc{AFDM_AF,
      title={A Unified Framework for Ambiguity Function Shaping and PAPR Control in AFDM Systems}, 
      author={Lingsheng Meng and Yong Liang Guan and Zilong Liu and Yirui Luo and Pingzhi Fan},
      year={2026},
      eprint={2604.22198},
      archivePrefix={arXiv},
      primaryClass={eess.SP},
      url={https://arxiv.org/abs/2604.22198}, 
}
\end{document}